\documentstyle[12pt,epsf]{article}
\begin{document}
\baselineskip 8mm
\sloppy
\thispagestyle{empty}

\vspace*{1cm}

\vspace*{1cm}
\begin{center}
{\Large\bf Spin and Chirality Orderings of Frustrated Magnets --- 
Stacked-Triangular
Antiferromagnets and Spin Glasses
}
\end{center}

\vspace*{2cm}

\begin{center}
{\large Hikaru Kawamura} \\
\bigskip

{\em Department of Earth and Space Science, Faculty of Science,\\
Osaka University, Toyonaka 560-0043, 
Japan}
\end{center}

\vspace{1.2cm}

\begin{center}
{\large\bf ABSTRACT}
\end{center}
\medskip

``Chirality'' is a multispin quantity representing the sense or the
handedness of the noncollinear spin structures induced by
spin frustration. 
Recent studies have revealed that
the chirality
often plays an important role in the
ordering of certain frustrated magnets. Here I 
take up two such examples, stacked-triangular antiferromagnets and
spin glasses, where the inherent chiral degree of freedom
affects underlying
physics and might lead to novel ordering phenomena.
The first topic is the criticality of the 
magnetic phase transition of vector ({\it i.e.\/},
$XY$ or Heisenberg) antiferromagnets on the three-dimensional 
stacked-triangular
lattice. The second topic is the nature of 
the spin-glass ordering. I will review the recent theoretical
and experimental works on these topics, with particular
emphasis on the important role played by the chirality.


\newpage

\noindent
\S 1. Introduction
\medskip

Frustration often gives rise to new interesting phenomena
in the magnetic ordering of spin systems. 
One interesting consequence of spin
frustration in vector spin systems might be the possible
appearance of ``chiral'' degrees of freedom.
``Chirality'' is a multispin quantity representing the sense or the
handedness of the noncollinear spin structures induced by
spin frustration. While the  chirality has long
been a familiar concept in the field of molecular chemistry, 
it was introduced into the field of magnetism first by 
Villain \cite{Villain}.
Recent studies have revealed that
the chirality
often plays a very important role in the
magnetic ordering of frustrated spin systems. 

In this article,  I will
take up two such examples, stacked-triangular antiferromagnets and
spin glasses, where the inherent chiral degree of freedom
leads to new interesting phenomena not encountered in unfrustrated
magnets.
The first topic is the criticality of the 
magnetic phase transition of vector ({\it i.e.\/},
$XY$ and Heisenberg) antiferromagnets on the three-dimensional (3D) 
stacked-triangular
lattice. Some time ago, 
the present author proposed that these frustrated 
triangular magnets,
{\it e.g.\/},
CsMnBr$_3$ and VBr$_2$, might
exhibit novel critical behavior, possibly 
lying in  a new ``chiral'' universality
class, distinct from the standard Wilson-Fisher universality
class \cite{HKtri1,HKtri2,HKtri3,HKtri4}. 
While this prediction have been supported by most of subsequent 
experiments, other theorists indicated that the transition was weakly
first order, 
and the subject still remains controversial \cite{HKtri5}.  
In this article,
I wish to review the
present status of the study, and discuss what to be done
in the future to further clarify the situation.

The second topic is the nature of the spin-glass ordering.
Our present understanding of the nature of the experimentally
observed spin-glass (SG) transition and of the SG ordered
state still remains unsatisfactory \cite{SGreview}. 
Although the chirality
has rarely been invoked in the 
standard scenario of the SG ordering, 
I with to explain here one promising  
scenario presented some time ago based on the 
spin-chirality decoupling-recoupling mechanism \cite{HKSG1}.
In this scenario, 
{\it chiral-glass order\/}, which is expected to occur
in a fully isotropic 3D
Heisenberg SG, 
plays an essential role. I examine 
some of the consequences of
this scenario, particularly 
in light of the  experimental results on canonical SG.
\bigskip
\bigskip

\noindent
\S 2. Chirality
\medskip

Frustration in vector spin systems often gives rise 
to the 
noncollinear or noncoplanar spin orderings. Such
canted spin structures generally give rise to the nontrivial
chiral degrees of freedom.
Chirality basically represents the handedness of such
noncollinear (noncoplanar) spin structures.
Two different types of chiralities have  often been discussed
in the literature: One is called a vector chirality and the other
a scalar chirality. 

\begin{figure}[ht]
\begin{center}
\noindent
\epsfxsize=0.63\textwidth
\epsfbox{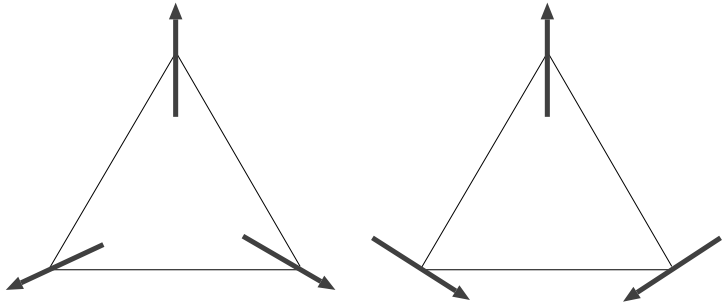}
\end{center}
\caption{Two chiral states in the ground-state spin
configurations of antiferromagnetically-coupled three vector spins
on a triangle. These chiral states are characterized by the
vector chirality.
}
\end{figure}

Chiral states representing the right- and left-handed configurations
are illustrated in Fig.1 for an example of three
antiferromagnetically coupled {\it XY\/} spins 
located at each corner of a triangle. The ground-state
spin configuration
is a well-known $120^\circ $ spin structure, in which each
{\it XY\/} spin on a plane 
makes an angle equal to $\pm 120^\circ$ with the neighboring
spins. One may define 
the chirality of the first type, the vector chirality,
via a vector product of the two neighboring 
spins, averaged over three spin pairs, by
\[
\kappa =\frac{2}{3\sqrt 3}\sum _{<ij>} 
\left [ \vec S_i\times \vec S_j\right ]_z.
\]
Evidently, the sign of $\kappa $ represents each of the two
chiral states,
{\it i.e.\/}, either a 
right-handed (clockwise) state for $\kappa >0$ or a 
left-handed (counterclockwise) state for $\kappa <0$.
In the case of {\it XY\/} spins considered here, the chirality
$\kappa $ is actually a
{\it pseudoscalar\/}: 
It remains invariant under global
$SO(2)$ proper spin rotations while it changes sign under
global $Z_2$ spin reflections. Hence, 
in order to transform one chiral state 
to the other, one needs to make
a global spin reflection. The
chiral order is closely related
to the spontaneous breaking of a discrete $Z_2$ spin-reflection
symmetry.

When the spins in Fig.1 are Heisenberg spins with
three components,
the ground-state spin configuration is again the coplanar
$120^\circ$ spin structure, whereas 
in contrast to the $XY$ case
there no longer
exists a discrete chiral degeneracy. This is because 
the apparently right-handed
chiral state can now be transformed into the apparently 
left-handed chiral state via a continuous
spin rotation of $\pi$ 
making use of the third dimension of the Heisenberg spin space. 
In this case,
one may define the vector chirality as an {\it axial vector\/} by
\[
\vec \kappa =\frac{2}{3\sqrt 3}\sum _{<ij>}\vec S_i\times \vec S_j.
\]

In the case of three-component Heisenberg spins,
there are occasions in which the ordered state spin configuration is
{\it noncoplanar\/} rather than coplanar.  Such a noncoplanar ordered
state is actually realized
in a Heisenberg SG. 
In fact, a noncoplanar state can sustain
a  discrete chiral degeneracy even in the case of Heisenberg spins,
which is characterized by  mutually opposite signs of the chirality
of the second type, the scalar chirality. It is a pseudoscalar
defined for three neighboring spins by,
\[
\chi =\vec S_i\cdot \vec S_{i+\delta} \times \vec S_{i+\delta '}.
\]
An illustrative example is given in Fig.2. 

\begin{figure}[ht]
\begin{center}
\noindent
\epsfxsize=0.63\textwidth
\epsfbox{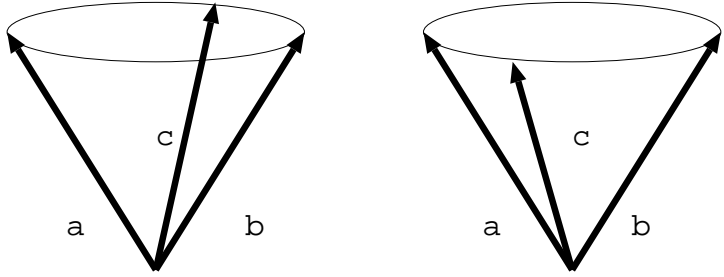}
\end{center}
\caption{Two chiral states associated with the noncoplanar
spin structure formed by three-component Heisenberg spins.
These chiral states are characterized by the
scalar chirality.
}
\end{figure}

These chiral degrees of freedom inherent to the noncollinear or
noncoplanar spin orderings often deeply affect the
ordering of frustrated magnets, 
leading to novel  phenomena not encountered in
conventional unfrustrated magnets, as we shall see in
the next two sections.

\bigskip
\bigskip
\bigskip
\bigskip

\noindent
\S 3. Critical properties of stacked-triangular antiferromagnets
\medskip

In this section, I wish to deal with the novel critical
behavior of {\it XY\/} and
Heisenberg antiferromagnets on the 3D stacked-triangular lattice.
The stacked-triangular
lattice consists of 2D triangular layers stacked in register
along the orthogonal direction directly on top of each other. 
There are several experimental realizations of the
stacked-triangular antiferromagnet with nontrivial chiral
degree of freedom. Examples are 
ABX$_3$-type compounds CsMnBr$_3$ and CsVBr$_3$, or vanadium
dihalides VCl$_2$ and  VBr$_2$, {\it etc\/}. Even Ising-like 
ABX$_3$-type compounds with an easy-axis-type anisotropy, 
such as CsNiCl$_3$, CsNiBr$_3$ and CsMnI$_3$, 
exhibit the chiral critical behavior under external fields higher than
a certain critical field.

The ordered-state spin configuration in these chiral 
stacked-triangular 
antiferromagnets
is a three-sublattice  $120^\circ$ spin structure in each 
triangular layer. 
As was illustrated
for the case of an isolated triangle,
such an ordered state 
possesses a nontrivial chiral degree of freedom. 
This leads to the appearance of
a new symmetry distinct from that of the standard unfrustrated
magnets, and
may give rise to magnetic phase  transitions
of new universality class. 
Indeed, the present author suggested 
the possible occurrence of such
a new universality class, chiral universality class,  on
the basis of 
a symmetry argument \cite{HKtri1,HKtri3}, Monte Carlo simulations 
\cite{HKtri1,HKtri4} and  
renormalization-group (RG) 
calculations including the $\epsilon =4-d$ and $1/n$
expansions \cite{HKtri2}. As usual,
we mean here by $n$ and $d$ the number of spin components and the spatial
dimensions, respectively, 
where the model is now extended to general $n$-vector spins
on a $d$-dimensional stacked-triangular lattice in which the
2D triangular layers are stacked in the remaining $d-2$ directions in a
hypercubic fashion. 
Theoretical analyses of 
the critical properties of $n$-vector
stacked-triangular
antiferromagnets performed by various authors, however, 
lead to often ambiguous, and sometimes
mutually conflicting results.
Below, I will refer to several points of special interest, leaving 
other details  to 
the review article \cite{HKtri5}.

The RG  $\epsilon =4-d$ expansion based on an 
appropriate Landau-Ginzburg-Wilson
Hamiltonian yields a new fixed point associated with 
the noncollinear criticality, distinct from the standard
Wilson-Fisher fixed point.
This new fixed point, which we call a chiral fixed point, 
exists stably 
in a certain region of the parameter space, at least for
sufficiently large $n$ meeting the condition, 
$n>21.8-23.4\epsilon +7.1\epsilon ^2+O(\epsilon ^3)$ \cite{HKtri2,Anto}.
For $n$ smaller than this critical value, the transition is likely
to be first order. 
The $1/n$ expansion result is also consistent with 
the $\epsilon =4-d$ expansion result, yielding 
a continuous transition with a set of
novel critical exponents  for any $2<d<4$ \cite{HKtri2}. 
The question still remaining is what happens at the physical point,
{\it i.e.\/}, $n=2$ or $3$ at $d=3$.
Two possibilities seem to exist:
(i) The critical value
$n_c(d=3)$ is smaller than, or equal to the physical value $n=2$
or 3.  If this is the case,  
a real system exhibits a continuous
transition of  new chiral universality. 
(ii) The critical value
$n_c(3)$ is greater than the physical value $n=2$ or 3. 
If this is the case,  a real system exhibits a first-order 
transition.

The present author 
originally suggested, partly based on an extrapolation of
the low-order $\epsilon $-expansion result, but mainly based on
the Monte Carlo observation, that  the case (i) was
a rather likely situation  \cite{HKtri2}.
By contrast,
some other theorists argued that the case (ii) was more likely:
For example,
on the basis of the Borel-Pad\'e resummation of the 
$\epsilon $-expansion, 
Antonenko {\it et al\/} 
estimated the critical value as $n_c(3)\simeq 3.39$
which turned out to be slightly above the physical value
$n=3$. Thus, the 
subject still remains controversial. 

Almost all Monte Carlo (MC) 
simulations so far made on the $XY$ ($n=2$)
and Heisenberg ($n=3$) stacked-triangular antiferromagnets yielded
a continuous transition characterized by the set of
novel exponents distinct from the standard $O(n)$ exponents 
\cite{HKtri4,Bhatta,Mailhot,Loison}.
We show in Table I  various critical
exponents and  specific-heat amplitude ratio as determined
by several MC simulations for the stacked-triangular Heisenberg ($n=3$)
antiferromagnet.
As is clear from  Table I, 
the MC values reported by different authors agree with each other, but
are inconsistent with the standard  Heisenberg values.
They also deviate considerably from the $O(4)$ or 
the mean-field tricritical 
values suggested in Ref.\cite{Azaria} 
on the bases of the RG $\epsilon =d-2$ expansion.

Since the simulations are performed for finite lattices, 
one cannot completely rule out the possibility that an
indication of a first-order transition eventually develops for still
larger lattices. A signature of such 
a first-order transition, however, 
has not been observed up to the linear size $L=60$.
Thus, the MC results so far obtained
are in favor of the chiral-universality scenario.
We note in passing that the numerical 
situation in the $XY$ ($n=2$) case is also
similar to 
the Heisenberg ($n=3$) case: See Ref.\cite{HKtri5} for details.

We have included in Table I 
the chirality exponents of the Heisenberg 
stacked-triangular antiferromagnet
as determined by MC, where $\beta _\kappa$ and $\gamma_\kappa$ refer
to the chiral long-range order 
and the conjugate chiral susceptibility exponents, respectively. The
RG analysis of Ref.\cite{HKtri2} 
indicated that the chirality was a new relevant
operator at the chiral fixed point. Although the chirality 
orders simultaneously with
the spin, characterized by 
the same chiral correlation-length exponent $\nu _\kappa $
as the spin correlation-length-exponent $\nu $,  
the chirality carries its own anomalous
dimension which is different from the spin anomalous dimension.
The chirality 
has a unique chiral crossover exponent $\phi _\kappa
=\beta_\kappa +\gamma_\kappa$, with satisfying 
the scaling relation $\alpha +2\beta _\kappa +\gamma _\kappa =2$.
These RG predictions were also supported by the 
MC results of Ref.\cite{HKtri2}.

Following the theoretical suggestion of 
a new chiral universality class, 
various experiments
have also been performed on stacked-triangular antiferromagnets
to test the prediction. The 
critical exponents and the specific-heat amplitude ratio
experimentally determined 
for the $XY$ ($n=2$) chiral magnets, {\it e.g.\/}, CsMnBr$_3$ 
\cite{Mason,Ajiro,Wang,Deutsch,Gaulin} and
CsNiCl$_3$ (CsMnI$_3$) above the critical (multicritical) 
field \cite{Beckman,
Enderle1,Enderle2,Weber},
are summarized in Table II. 
All experiments  reported that the transition was
continuous. 
Several theoretical values are also shown for comparison, 
including the $n=2$ chiral values of Ref.\cite{HKtri4}.
As can be seen from Table II, these experimental
values are consistent with each other, and are in
favor of the 
chiral-universality scenario. We note that the experimental 
situation in the Heisenberg  case is also
similar to 
the $XY$ case: See Refs.\cite{HKtri5} and \cite{trireview}
for details. 

At this point, I wish to add a few
comments. First, even when the chiral
fixed point is stable with a finite basin of attraction in
the parameter space,
it is still entirely possible that
some systems possessing the full chiral symmetry exhibit a first-order
transition. This occurs 
only if the initial Hamiltonian characterizing the system
happens to lie outside the basin of the chiral fixed point.
Here the difference between the two classes of systems, one
showing a continuous chiral transition and the other 
a first-order transition, 
is not of symmetry origin, but arises simply
from the nonuniversal details of the microscopic parameters.
Possible examples of chiral systems showing
a first-order transition might be 
certain matrix models with the rigid noncollinear
structures. Indeed, the initial RG Hamiltonians describing
these matrix models are likely to be located in the runaway regime
of the RG flow, outside the stability domain of the chiral fixed 
point \cite{HKtri5}. Conversely, 
this observation means that, even if one finds a few
chiral systems exhibiting a first-order transition, it does not 
immediately exclude
the possibility that still another class of systems with the
same chiral symmetry exhibit a 
continuous transition belonging to chiral universality.

Second, the $\epsilon =4-d$ expansion suggests that
the critical value of $n_c(d=3)$  may lie very
close to the physical value $n=2$ or 3. 
If $n_c(3)$ lies above the physical 
value only very slightly, 
there is no stable fixed point in the strict sense (the chiral
fixed point  becomes slightly
complex-valued). Nevertheless,  the associated RG flow behaves 
for a long period of RG iterations as if there were a 
stable fixed point. Thus, a ``shadow'' of the complex-valued chiral fixed
point in the  real plane attracts the RG flow up to a certain scale,
which could be very long, but, eventually, the flow escapes away from
such a pseudo-fixed point showing a runaway characteristic 
of a first-order
transition. Physically, this means that the system exhibits a rather
well-defined critical behavior governed by
the (slightly 
complex-valued) chiral fixed point for a wide range of temperature,
but eventually, the deviation from
such a critical behavior sets in for sufficiently close to 
$T_c$, and the system exhibits a weak first-order transition. We 
emphasize
here that, as long as $n_c(3)$ lies very close to the physical value,
the situation would virtually be 
indistinguishable  for all practical purposes
from the case of a truly
stable chiral fixed point,
in the sense that almost universal novel
critical behavior
is observed  for a variety of chiral systems  
in a wide temperature range up to
very close to $T_c$ \cite{HKtri5}.

Third, besides the $\epsilon =4-d$ and $1/n$ expansions, 
there is another perturbative RG scheme called the $\epsilon =d-2$
expansion \cite{Azaria,HKtri6}. 
This method applied to the $n=3$ chiral systems yielded a
continuous transition characterized by 
the standard $O(4)$ criticality. Meanwhile,
as shown above, such an $O(4)$ behavior has not been observed 
either
numerically nor experimentally
in the magnetic ordering of stacked-triangular antiferromagnets. 
The possible reason of this 
failure of the $\epsilon =d-2$
expansion 
was discussed
by several authors \cite{HKtri6,KZ,DJ,DD}. 
One reason might be 
that the nonperturbative
effects associated with the topological defects,
which might play a vitally important role in the phase transition of
chiral systems,
was completely neglected
there \cite{HKtri6,KZ}.
In particular, if the $n=3$ chiral system in $d=2$ dimensions exhibits 
a pure topological
phase transition at a finite temperature driven by the dissociation
of the $Z_2$ vortices, as first suggested in Ref.\cite{KM} and 
numerically supported in Refs.\cite{KK,Southern,S}, 
the $\epsilon =d-2$ expansion should completely 
fail {\it even at and near two dimensions\/}. 
Then, to construct a phase diagram of the chiral system in the entire
$d-n$ plane ($2\leq d\leq 4$) remains as an open question.

Before concluding this section of stacked-triangular
antiferromagnets, I wish to briefly 
discuss the future problems and challenges, mainly on 
experimental side.

i) The first obvious direction is to perform high-precision 
experiments (or numerical simulations) to see whether the chiral
transition is either continuous or first order, and to precisely
determine the associated critical exponents and amplitude ratio. 
In particular, 
if the chiral transition is really of weakly
first order, it should eventually
show up in high-precision experiments as a finite latent heat or 
a finite
discontinuity in certain physical quantities. In doing this, one has to
choose appropriate materials characterized by the full chiral symmetry. 
Otherwise,
a weak but nonzero perturbation of lower symmetry,
{\it e.g.\/}, the ones arising from the lattice distortion as in the
cases of RbMnBr$_3$ and CsCuCl$_3$, 
would
cause a crossover to the non-chiral behavior in a close 
vicinity of $T_c$,
irrespective of the eventual stability of the chiral fixed point.

ii) The other direction is to further examine the underlying 
{\it universality\/} 
of the novel chiral critical behavior, {\it i.e.\/}, to examine the
critical properties from one  material to the other.
If the novel chiral criticality is to be observed in various chiral
magnets in common, it should be a  good experimental proof that
there indeed exists a chiral fixed point behind the phenomena.

iii) Another interesting point, closely 
related to the point ii) above, is
to examine the critical properties of helical magnets (spiral magnets),
which is expected to lie in the same universality class as the 
stacked-triangular antiferromagnets. So far, the critical
properties of helimagnets with nontrivial chiral degrees of freedom
have been studied almost exclusively for rare-earth metals,
Ho, Dy and Tb. However, the situation for rare-earth helimagnets
has remained confused for years now \cite{HKtri5}.
Apparently, there
are serious problems associated with the
long-range nature of the RKKY interaction, or with the possible 
contribution of the second
length scale arising from the
disorder-containing skin part of the sample. One way  to avoid these
complications might be to make measurements on {\it insulating\/} 
helimagnets
such as VF$_2$ and $\beta -$MnO$_2$, and I urge experimentalists to try
such experiments.

iv) Since the chirality is a quantity characteristic of the novel chiral 
criticality,  a direct experimental observation of the chirality
would be very useful.
Recently, a significant progress in this direction was made  
by Plakty {\it et al\/},
who determined by using elaborate experimental technique
the critical behavior of the chirality for
the stacked-triangular antiferromagnets CsMnBr$_3$ and CsNiCl$_3$ 
\cite{Plakhty}. The
determined chirality exponents
are compared with the prediction of the 
chiral universality quite successfully.

We finally note that, while the chiral degrees of freedom 
inherent to stacked-triangular antiferromagnets and other chiral
magnets give rise to
the novel critical behavior, quite possibly lying in a new
universality class, the chirality itself behaves  
as a composite operator
of the order parameter of the transition, the spin.
In other words, the spin and the chirality are not decoupled here, 
and
there is only one diverging length scale in the
transition. Under certain circumstances, however, even this property
could change, as we shall see in the next section on the
SG ordering.
\bigskip
\bigskip

\noindent
\S 4. Ordering in spin glasses 
\medskip

Spin glasses are the type of random magnets in which both ferromagnetic 
and antiferromagnetic interactions coexist and compete, 
thereby giving rise to the effects of
frustration and quenched randomness. 
Since the experimental discovery of a sharp transition-like
phenomenon in typical SG magnets, particularly in the so-called 
canonical SG that are
dilute metallic alloys such as AuFe and CuMn, 
their ordering properties  have been 
studied quite extensively both experimentally and theoretically
\cite{SGreview}.
Nevertheless, the true nature of the SG 
ordering
still remains elusive, and is under hot debate. 

Since we now have fairly 
convincing experimental
evidences that typical SG magnets exhibit 
an equilibrium  phase transition 
at a finite temperature, next obvious question is what is
the true nature of the experimentally observed SG
transition and of the low-temperature SG state.
We have not yet reached the full understanding of this question.
In theoretical studies of the ordering of SG, {\it e.g.\/}, of
the critical
properties of the SG transition or of the 
issue of whether the SG state exhibits a spontaneous
replica-symmetry breaking (RSB),
a simple Ising model with short-range interaction,
the so-called Edwards-Anderson (EA) model, 
has widely been used as a ``realistic'' SG
model. 
One should bear in mind, however, that 
the magnetic interactions in  many SG
materials are nearly isotropic, being 
rather well described by an isotropic
Heisenberg model, in the sense that the magnetic anisotropy is
considerably weaker than the exchange interaction.  
In apparent contrast to experiments,
numerical simulations have indicated
that the standard SG order 
occurs only at zero temperature in the 3D
Heisenberg SG \cite{OYS,Matsubara,HKSG1,HKSG2,HK}. 
Although the magnetic anisotropy inherent 
to real materials
is often invoked to explain this apparent discrepancy with 
experiments,
it still remains puzzling that no
detectable sign of
Heisenberg-to-Ising crossover has been observed in experiments
which is usually expected to occur if the observed
finite-temperature transition 
is caused by the weak magnetic anisotropy\cite{SGreview,OYS,Gingras}. 

In this section, I wish to explain 
one scenario of the experimental SG ordering presented 
some time ago by the present
author, 
aimed at solving  this apparent puzzle\cite{HKSG1}. 
In this scenario, which may be termed as a spin-chirality
decoupling-recoupling scenario, a scalar chirality
introduced in \S 2 plays an 
important role.
This scenario consists of two parts: The fist part is  
``spin-chirality
separation'' or ``decoupling'' for the fully isotropic system, and the
second part is the ``spin-chirality mixing'' or ``recoupling'' due to
random magnetic anisotropy. 

Recent numerical simulations indicate that, 
in a fully isotropic Heisenberg SG
in 3D, the chirality 
is ``separated'' on long length and time scales from the spin, 
giving rise to a novel chiral-glass
ordered phase in which
only the scalar chiralities are ordered  in a spatially 
random manner 
with keeping the spin 
paramagnetic \cite{HKSG2,HK}. 
In other words, in a 3D isotropic Heisenberg SG,
the chiral-glass order precedes the SG order, 
$T_{{\rm CG}}>T_{{\rm SG}}$, the latter probably taking place
only at
$T=T_{{\rm SG}}=0$. 
Furthermore, a recent numerical simulation has revealed that
the nature of the chiral-glass transition and of the chiral-glass
ordered state might differ in some essential points from that of the
standard Ising SG \cite{HK}. First, the
critical properties of the chiral-glass transition, 
characterized
by the exponents 
$\beta _{{\rm CG}}\simeq 1.1,\ \gamma _{{\rm CG}}\simeq 1.5,
\nu _{{\rm CG}}\simeq 1.2$ and $\eta _{{\rm CG}}\simeq 0.8$ 
{\it etc.\/}, 
differ from those of the standard 3D Ising spin glass, characterized by
the exponents
$\beta \simeq 0.55,\ \gamma \simeq 4.0,
\nu \simeq 1.8$ and $\eta \simeq -0.35$, {\it etc\/} 
\cite{KY,Marinari,PC}.
Second, 
the chiral-glass ordered state is 
likely to exhibit a novel type of replica-symmetry breaking (RSB)
{\it with a one-step-like character\/}, in contrast to the case of
the 3D Ising SG.

Of course, the experimentally observed SG transition is associated
with the freezing of the spin itself: 
Experimentally, the scalar chirality has not been
directly measurable in SG so far. Since the chiral-glass order expected
in a 3D isotropic Heisenberg SG does not accompany the 
spin order,
it cannot immediately be  connected with the experimental SG ordering. 
At this point, 
we come to the second part of the scenario, ``spin-chirality
recoupling'' or ``mixing''. Here, one assumes that
the weak random magnetic anisotropy 
inherent to
real SG materials ''mixes'' or ''recouples'' the spin and the
chirality, and the chiral-glass transition, which is hidden in the
chiral sector in the absence of magnetic anisotropy, is now
''revealed''  in the spin sector. As long as
the anisotropy is sufficiently weak, the nature of the SG transition 
and of the SG
ordered state should be governed by that of the 
chiral-glass transition and of the chiral-glass ordered state
in the fully isotropic system. 
Note that it is {\it not\/} governed by the the SG transition
and the SG ordered state of the fully isotropic system, 
since the spin has been separated from the chirality there.

One can see such a spin-chirality mixing  from a simple symmetry
argument.
The random magnetic anisotropy generally reduces the global
symmetry of the Hamiltonian from  $O(3)=Z_2\times SO(3)$ to
only {\it chiral\/} 
$Z_2$ associated with the global
spin inversion $\vec S_i\rightarrow -\vec S_i$.
Note that the spin 
inversion flips the scalar chirality 
$\chi _i\rightarrow -\chi _i$, since $\chi _i$ is cubic in spins.
As mentioned,
in the fully isotropic case, the spin-chirality separation is 
expected to occur. There, upon renormalization, 
the Hamiltonian is asymptotically decoupled
into the $Z_2$-symmetric
chiral part and the $SO(3)$-symmetric  part, 
where  the $Z_2$ chiral part is subject to
a finite-temperature chiral-glass
transition, while the $SO(3)$ part remains
disordered keeping the Heisenberg spin paramagnetic even in the
chiral-glass ordered state. Suppose that a small amount of
random magnetic anisotropy is added to the system. 
While the random magnetic anisotropy  energetically breaks the
$SO(3)$ symmetry, 
the chiral-glass transition associated with the chiral $Z_2$ symmetry
would persist essentially unchanged, 
since the random anisotropy preserves the $Z_2$ chiral symmetry 
and
the $Z_2$ chiral part has
already been separated from the $SO(3)$ part. In contrast to the fully
isotropic case,
once the chiral-glass transition
takes place and the $Z_2$  symmetry is spontaneously broken,
the spin is no 
longer allowed to remain paramagnetic since the system now 
does not possess the $SO(3)$ symmetry.
Thus, the spin is now ``forced to order'' in the chiral-glass
state via the  effective coupling
between the spin and the chirality generated by the random magnetic
anisotropy.

Based on this physical picture, one can derive various interesting
predictions for the real Heisenberg-like 
SG ordering. For example, the SG
transition temperature in the presence of random magnetic anisotropy
of magnitude $D$ is expected to behave for small $D$ as 
%
\[
T_{{\rm SG}}(D)\approx T_{{\rm CG}}(0)\left [1+c\left (\frac{D}
{k_BT_{{\rm CG}}(0)}\right )^2+\cdots \right ],
\]
where $T_{{\rm CG}}(0)$ is the chiral-glass transition temperature
of the fully
isotropic system, and $c$
is a numerical constant.
Here note that $T_{{\rm CG}}(0)$ is a quantity of $O(J)$, 
not of $O(D)$. Hence,  
in the $D\rightarrow 0$ limit,  $T_{{\rm SG}}(D)$ tends to a 
finite value of $O(J)$ essentially
in a regular way.
The reason why the $T_{{\rm SG}}(D)$ in the
$D\rightarrow 0$ limit tends to $T_{{\rm CG}}(0)$ associated with the
$Z_2$-symmetry breaking, 
not to $T_{{\rm SG}}(0)(<T_{{\rm CG}}(0))$ associated with the
$SO(3)$-symmetry breaking,  
is because
the symmetry spontaneously
broken at the SG transition under finite $D$
is the 
chiral $Z_2$ symmetry (spin-inversion symmetry),
not being the continuous $SO(3)$ symmetry.

The nonlinear susceptibility for sufficiently weak $D$
diverges toward $T=T_{{\rm SG}}(D)$ as
\[
\chi_{nl}(D) \approx D^4t^{-\gamma _{{\rm CG}}} + [{\rm nondiverging\ 
term}],
\]
where $t$ is a reduced temperature $t\equiv |(T-T_{{\rm SG}}(D))/
T_{{\rm SG}}(D)|$. 
The susceptibility exponent here is nothing but the chiral-glass
exponent in the fully isotropic system 
$\gamma =\gamma _{{\rm CG}}\simeq 1.5$. 
The above equation indicates that
the standard crossover behavior in the exponent does not occur 
even for weak $D$. In contrast, 
the {\it amplitude\/} of the leading singularity
depends on $D$, vanishing in the isotropic limit. In this way, 
the present chirality scenario gives a natural explanation of the
abovementioned puzzles about experiments, {\it i.e.\/}, why
the expected Heisenberg-to-Ising
crossover has not been observed experimentally, or why the SG
transition occurs at a finite temperature of order $J$ even in a
nearly isotropic Heisenberg-like SG with weak anisotropy.

According to the present chirality scenario, the
critical properties of the SG transition of
real Heisenberg-like SG should be  the same as
those of the chiral-glass transition of an isotropic Heisenberg
SG, {\it i.e.\/}, 
$\beta \simeq 1.1,\ \gamma \simeq 1.5,\ \nu \simeq 1.2$ and
$\eta \simeq 0.8$.  These exponent values, though quite different
from the values of the 3D Ising SG, happen to be rather close
to the experimental values  for canonical SG.
In Table III, I summarize the SG critical exponents experimentally
determined for canonical SG 
such as AgMn, CuMn {\it etc\/}.
One sees that the exponents determined by various authors
\cite{Levy,Courtenary,Bouchiat,Simpson,Coles,Vincent}
come  close to each other,
yielding the values $\beta 
\simeq 1$, $\gamma \simeq 2$, $\nu \simeq 1.3$
and $\eta \simeq 0.5$. 
In Table III, we also give the corresponding exponent
values of the 3D Ising SG determined by recent extensive
numerical simulations \cite{KY,Marinari,PC}.  
Evidently, the experimental
values for canonical SG deviate considerably from the 3D Ising 
values, while they come rather close to the
chiral-glass values quoted above, 
giving some support to the chirality scenario.

It should be noticed  that one cannot ascribe
the cause of the observed large 
discrepancy between the experimental and the
3D Ising exponent values  to
the long-range nature of the
RKKY interaction inherent to canonical SG.
Although the RKKY interaction is expected to be a relevant
perturbation in the spin ordering of 3D Heisenberg SG \cite{BMY},
a scaling theory suggests that the 3D Heisenberg
SG with the RKKY interaction lies 
just at its lower critical dimension (LCD) \cite{BMY}. Generally, 
the system at its LCD is known to exhibit either a zero-temperature
transition with an exponentially diverging correlation length,
or a finite-temperature Kosterlitz-Thouless-type transition
without a finite long-range order but with
an exponentially diverging correlation length. In either case, $\nu $
should be infinite, which is hardly consistent with the experimental
result for canonical SG, $\nu \simeq 1.3$. 
Furthermore, the non-Ising exponents close to the
ones obtained for canonical SG are also observed in an insulating
Heisenberg-like SG with short-range interaction, CdCrInS 
\cite{Vincent}: See Table III. These observations indicate that the 
cause of the non-Ising exponents experimentally 
observed in canonical SG
cannot simply be ascribed to the RKKY interaction.

Thus, the chirality
scenario turns out to naturally explain not only the origin of the
experimentally observed SG exponents, 
but also the reason why the expected
crossover has not been observed experimentally, or why the SG
transition occurs at a finite temperature of $O(J)$ even in a
nearly isotropic Heisenberg-like SG system.
In fact, this scenario could also explain some other features where
the standard theory met some difficulty, such as the
problem of magnetic phase diagram \cite{HKunpub}. 
It also yields some new predictions, {\it e.g.\/}, that the
SG ordered state of real Heisenberg-like SG
should exhibit a peculiar type of one-step-like
RSB.  Further details of the chirality scenario and its outcome
will be given elsewhere \cite{HKunpub}. 
\bigskip
\bigskip

\noindent
\S 5. Concluding remark
\medskip

A brief review has been given on the two topics in 
the ordering of frustrated
magnets where the chirality plays an important role.
One is the criticality of stacked-triangular antiferromagnets
and the other is the ordering of spin glasses.
In both cases, the chirality gives rise to new interesting phenomena
not encountered in standard unfrustrated magnets, such as
``chiral universality'' or ``spin-chirality separation''. 
While further works are required to settle the issues discussed in
this article, it is already clear that
frustration  indeed offers 
a unique stage where a variety of
new interesting ordering phenomena take place.

The author  is particularly indebted  to Dr. K. Hukushima
for useful discussion and collaboration in a work cited
in \S 4.

\bigskip\bigskip

\noindent
{\it Note added to \S 3\/}: Very recently, Pelissetto, Rossi
and Vicari performed a RG loop expansion at $d=3$ up to
six loops [A. Pelissetto, P. Rossi and E. Vicari, 
Phys. Rev. B{\bf 63}, R140414 (2001)]. 
By determining the large-order behavior of the
series and making a resummation, these authors found a stable
chiral fixed point for both cases of $n=2$ and 3
characterized by the exponents which were in good
agreement with experiments and MC. Note that
the result is in sharp contrast to
the previous lower-order (three-loops) result where
no stable fixed was found.

\newpage
\fussy

\newpage

\begin{tabular}{l|l l l l|l l l} \hline\hline
  & Kawamura & Bhattacharya  
 & Mailhot  & Loison  
 & Heisen- & O(4)  & MF tri-
\\ 
Ref. & & {\it et al\/} & {\it et al\/} & {\it et al\/} 
 & berg & 
 & critical
\\ 
 & \cite{HKtri4} & \cite{Bhatta} & \cite{Mailhot} & 
 \cite{Loison} & & & 
\\ \hline
$T_c/J$ & 0.958(4) & 0.9576(2) & 0.9577(2) & - & & & 
\\
$\alpha$ & 0.24(8) & - & - & - & -0.116 & -0.22 & 0.5
\\
$\beta$ & 0.30(2) & 0.289(10) & 0.285(11) & 0.28(2) & 0.36 & 0.39 & 0.25
\\
$\gamma$ & 1.17(7) & 1.176(20) & 1.185(3) & 1.25(3) & 1.387 & 1.47 & 1
\\
$\nu$ & 0.59(2) & 0.585(9) & 0.586(8) & 0.59(1) & 0.705 & 0.74 & 0.5
\\
$A^+/A^-$ & 0.54(20) & - & - & - & 1.36 & - & 0
\\ \hline
$\beta_\kappa$ & 0.55(4) & - & 0.50(2) & - & - & - & 0.5
\\
$\gamma_\kappa$ & 0.72(8) & - & 0.82(4) & - & - & - & 0.5
\\
$\nu _\kappa$ & 0.60(3) & - & 0.608(12) & - & - & - & 0.5
\\ \hline\hline
\end{tabular}

\bigskip\medskip\noindent
{\bf Table I:} 
Critical exponents, specific-heat amplitude ratio and
transition temperature as determined by several MC simulations
for the stacked-triangular Heisenberg antiferromagnet 
with the equal strengths of the inter- and intra-plane
couplings $J$. The corresponding values of the standard
Heisenberg, $O(4)$ and mean-field (MF) tricritical universality
classes are also shown.
\bigskip\bigskip\bigskip

\begin{tabular}{l|l l l l l} \hline\hline
 & $\alpha $ & $\beta $ & $\gamma $ & $\nu $ & $A^+/A^-$
\\ \hline
CsMnBr$_3$ & 0.39(9)\cite{Wang} & 0.22(2)\cite{Mason} & 
 1.10(5)\cite{Ajiro}  & 
 0.57(3)\cite{Ajiro}  & 0.19(19)\cite{Wang} 
\\
 & 0.40(5)\cite{Deutsch}  & 0.25(1)\cite{Ajiro} & 
 1.01(8)\cite{Mason} & 
 0.54(3)\cite{Mason} & 0.32(20)\cite{Deutsch} 
\\
 & & 0.24(2)\cite{Gaulin}  &  &  & 
\\
CsNiCl$_3$ & 0.37(8)\cite{Beckman} & 0.243(5)\cite{Enderle2} & 
 - & - & 
 0.30(11)\cite{Beckman} 
\\
\ ($H>H_m$) & 0.342(5)\cite{Enderle1} & & & & 
\\
CsMnI$_3$ & 0.34(6)\cite{Weber} & - & - & - & 0.31(8)\cite{Enderle1} 
\\
\ ($H>H_m$) & & & & & 
\\ \hline
{\it XY\/} & -0.008 & 0.35 & 1.316 & 0.669 & 0.99
\\
$n=2$ chiral\cite{HKtri4} & 0.34(6) & 0.253(10) & 1.13(5) & 
 0.54(2) & 0.36(20)
\\
MF tricritical & 0.5 & 0.25 & 1.0 & 0.5 & 0 
\\ \hline\hline
\end{tabular}

\bigskip\medskip\noindent
{\bf Table II:} 
Critical exponents and specific-heat amplitude ratio
determined by the experiments on several
stacked-triangular {\it XY\/} antiferromagnets. The corresponding 
values given by several theories are also shown. 

\newpage

\begin{tabular}{l|l|l| l l l l} \hline\hline
type & material & Ref. & $\beta$ & $\gamma$ & $\nu$ &
 $\eta $
\\ \hline
canonical SG & CuMn,AgMn & \cite{Courtenary} & 1.0(1) & 2.2(1) & $\sim$ 1.4 
 & $\sim$ 0.4
\\
 & AgMn & \cite{Bouchiat} & 1.0(1) & 2.2(2) & $\sim$ 1.4 & $\sim$ 0.4
\\
 & AgMn & \cite{Levy} & 0.9(2) & 2.1(1) & $\sim$ 1.3 & $\sim$ 0.4
\\ 
 & CuAlMn & \cite{Simpson} & $\sim$ 1.0 & $\sim$ 1.9 & $\sim$ 1.3 & $\sim$ 0.5
\\
 & PdMn & \cite{Coles} & 0.90(15) & 2.0(2) & $\sim$ 1.3 & $\sim$ 0.4
\\ \hline
 insulating SG & CdCrInS & \cite{Vincent} & 0.75(10) & 2.3(4) & $\sim$ 1.3 & 
 $\sim$ 0.2
\\ \hline\hline
3D Ising SG & $\pm J$ & \cite{KY} & $\sim$ 0.55 & $\sim$ 4.0 & 1.7(3) & 
 -0.35(5)
\\
\ (MC) & Gaussian & \cite{Marinari} & $\sim$ 0.64 & $\sim$ 4.7 & 2.0(1.5) & 
 -0.36(6) 
\\
  & $\pm J$ & \cite{PC} & $\sim$ 0.65 & 4.1(5) &  1.8(2) & -0.26(4)
\\ \hline
chiral glass & & \cite{HK} & 1.1(1) & 1.5(3) & $\sim 1.2$ & $\sim 0.8$ 
\\ \hline\hline
\end{tabular}

\bigskip\medskip\noindent
{\bf Table III} 
Critical exponents of canonical SG (Heisenberg-like 
metallic SG) and of Heisenberg-like
insulating SG, compared with the corresponding
exponents of the
3D short-range Ising SG (EA model) and those of the
chiral-glass. Note that the standard scaling relations have been used
here to reproduce the full set of exponents from the values reported
in the original references.


%




\begin{thebibliography}{99}


\bibitem{Villain} J. Villain, J. Phys. C{\bf 9}, 4793 (1977).

\bibitem{HKtri1} H. Kawamura, J. Phys. Soc. Jpn. {\bf 54}, 3220 (1985);
{\bf 55}, 2095 (1986); {\bf 56}, 474 (1986); {\bf 58}, 584 (1989).

\bibitem{HKtri2} H. Kawamura, Phys. Rev. {\bf B38}, 4916 (1988); 
{\bf B42}, 2610 (1990) [E]; J. Phys. Soc. Jpn. {\bf 59}, 2305 (1990). 

\bibitem{HKtri3} H. Kawamura, J. Appl.  Phys. {\bf 63}, 3086 (1988).

\bibitem{HKtri4} H. Kawamura, J. Phys. Soc. Jpn. {\bf 61}, 1299 (1992).

\bibitem{HKtri5} H. Kawamura, J. Phys. Condens. Matter {\bf 10}, 4707 
(1998).

\bibitem{SGreview}
For reviews on spin glasses, see {\it e. g., }
(a) K. Binder and A. P. Young, Rev. Mod. Phys. {\bf 58}, 801 (1986); 
(b) K. H. Fischer and J. A. Hertz, {\it Spin Glasses} Cambridge 
University Press (1991); 
(c) J. A. Mydosh, {\it Spin Glasses} Taylor \& Francis (1993); 
(d) A. P. Young ({\it ed.\/}), 
{\it Spin glasses and random fields} World Scientific,
Singapore (1997). 

\bibitem{HKSG1}
H. Kawamura, Phys. Rev. Lett.  {\bf 68}, 3785 (1992);
Int. Jour. Mod. Phys. {\bf 7}, 345 (1996).

\bibitem{Anto}
S.A. Antonenko, A.I. Sokolov and K.B. Varnashev, Phys. Lett.  
{\bf 208}A, 3785 (1995).

\bibitem{Bhatta}
T. Bhattcharya, A. Billoire, F. Delduc and Th. Jolicoeur,
J. Physique {\bf 4}, 181 (1994).

\bibitem{Mailhot}
A. Mailhot, M.L. Plumer and A. Caille, Phys. Rev. B{\bf 50}, 6854 (1994).

\bibitem{Loison}
D. Loison and H.T. Diep, Phys. Rev. B{\bf 50}, 16453 (1994).

\bibitem{Azaria}
P. Azaria, B. Delamotte and Th. Jolicoeur, Phys. Rev. Lett.
{\bf 64}, 3175 (1990).

\bibitem{Mason} T.E. Mason, M.F. Collins and B.D. Gaulin: 
J.  Phys. C{\bf 20}, L945 (1987);
Phys. Rev. B{\bf 39}, 586 (1989).

\bibitem{Ajiro} Y. Ajiro, T. Nakashima, Y. Unno, H. Kadowaki, M. Mekata
and N. Achiwa, J. Phys. Soc. Jpn. {\bf 57} (1988) 2648;
H. Kadowaki, S.M. Shapiro, T. Inami and Y. Ajiro,
J. Phys. Soc. Jpn. {\bf 57} (1988) 2640.

\bibitem{Gaulin} B.D. Gaulin, 
T.E. Mason, M.F. Collins and J.Z. Larese, Phys. Rev. Lett. 
{\bf 62}, 1380 (1989).

\bibitem{Wang} J. Wang, D.P. Belanger and B.D. Gaulin,
Phys. Rev. Lett. {\bf 66}, 3195 (1991).

\bibitem{Deutsch} R. Deutschmann, H.v. L\"ohneysen, 
J. Wosnitza, R.K. Kremer, Europhys. Lett. {\bf 17}, 637 (1992).

\bibitem{Beckman} D. Beckman, J. Wosnitza and H. von L\"ohneysen, Phys.
Rev. Lett. {\bf 71}, 2829 (1993).

\bibitem{Enderle1} M. Enderle, G. Furtuna and M. Steiner,
J. Phys. Condens. Matter {\bf 6}, L385 (1994).

\bibitem{Enderle2} M. Enderle, R. Schneider, Y. Matsuoka and 
K. Kakurai, 
Physica B{\bf 234-236}, 554 (1997).

\bibitem{Weber} H. Weber, D. Beckmann, J. Wosnitza 
and H. von L\"ohneysen, Int. J. Mod. Phys. 
B{\bf 9}, 1387 (1995).

\bibitem{trireview}
M.F. Collins and O.A. Petrenko,
Can. J. Phys. {\bf 75}, 605 (1997).

\bibitem{HKtri6} H. Kawamura, J. Phys. Soc. Jpn. {\bf 60}, 1839 (1991).

\bibitem{KZ} H. Kunz and G. Zumbach, 
J. Phys. A{\bf 26}, 3121 (1993); G. Zumbach, 
Nucl. Phys. B{\bf 435}, 753 (1995).

\bibitem{DD}
A. Dobry and H.T. Diep, Phys. Rev. B{\bf 51}, 6731 (1995).

\bibitem{DJ}
F. David and Th. Jolicoeur,
Phys. Rev. Lett. {\bf 76}, 3148 (1996).

\bibitem{KM} H. Kawamura and S. Miyashita, 
J. Phys. Soc. Jpn. {\bf 53}, 4138 (1985).

\bibitem{KK}
H. Kawamura and M, Kikuchi, Rev. B{\bf 47}, 1134 (1993).

\bibitem{Southern}
B.W. Southern and H-J Xu, Phys. Rev. B{\bf 52}, R3836 (1995).

\bibitem{S}
W. Stephan and B.W. Southern, Phys. Rev. B{\bf 61}, 11514 (2000); talk in this
wokshop (to appear in Can. J. Phys.).

\bibitem{Plakhty}
V.P. Plakhty, 
talk in this workshop (to appear in Can. J. Phys.); S.V. Maleyev, 
V.P. Plakhty, O.P. Smirnov, J. Wosnitza, D. Visser, R.K. Kremer and J. Kulda,
J. Phys. Condens. Matter {\bf 10}, 951 (1998).

\bibitem{OYS}
J. A. Olive, A. P. Young and D. Sherrington, Phys. Rev. B{\bf 34},
6341 (1986).

\bibitem{Gingras}
We note that, 
numerically and in 2D, the expected
crossover to an Ising-like behavior has been observed
for the $T=0$ transition of the 2D Heisenberg-like SG with weak random magnetic
anisotropy:
See M.J.P. Gingras, Phys. Rev. Lett. {\bf 71},
1637 (1993).

\bibitem{Matsubara}
F. Matsubara, T. Iyota, and S. Inawashiro, Phys. Rev. Lett. {\bf 67},
1458 (1991).

\bibitem{HKSG2}
H. Kawamura, Phys. Rev. Lett. {\bf 80}, 5421 (1998).

\bibitem{HK}
K. Hukushima and H. Kawamura, Phys. Rev.  E{\bf 61}, R1008 (2000);
H. Kawamura and K. Hukushima, Int. J. Mod. Phys. C{\bf 10}, 1471
(1999).

\bibitem{Levy}
L.P. L\'evy and A.T. Ogielsky, Phys. Rev. Lett. {\bf 57}, 3288 (1986).

\bibitem{Courtenary}
N. de Courtenary, H. Bouchiat, H. Hurdequint and A. Fert,
J. Physique {\bf 47}, 1507 (1986).

\bibitem{Bouchiat} 
H. Bouchiat, J. Physique {\bf 47}, 71 (1986).

\bibitem{Simpson}
M. Simpson, J. Phys. F{\bf 9}, 1377 (1979).

\bibitem{Coles}
B.R. Coles and G. Williams, J. Phys. F{\bf 18}, 1279 (1988).

\bibitem{Vincent}
E. Vincent, J. Hammann and M. Alba,
Solid State Comm. {\bf 58}, 57 (1986); E. Vincent and J. Hammann,
J. Phys. C{\bf 20}, 2659 (1987).

\bibitem{KY}
N. Kawashima and A.P. Young, Phys. Rev. B{\bf 53}, 484 (1996).

\bibitem{Marinari}
E. Marinari, G. Parisi and J.J. Ruiz-Lorezo,  
Phys. Rev. B{\bf 58}, 14852 (1998).

\bibitem{PC}
M. Palassini and S. Caracciolo, 
Phys. Rev. Lett. {\bf 58}, 14852 (1998).

\bibitem{BMY}
A.J. Bray, M.A. Moore and A.P. Young, Phys. Rev. Lett. {\bf 56}, 
2641 (1986).

\bibitem{HKunpub}
H. Kawamura, unpublished.

%


\end{thebibliography}
\end{document}